# Constituent-Quark Model and New Particles

David Akers*

*Lockheed Martin Corporation, Dept. 6F2P, Bldg. 660, Mail Zone 6620,
1011 Lockheed Way, Palmdale, CA 93599*
*Email address: David.Akers@lmco.com

An elementary constituent-quark (CQ) model by Mac Gregor is reviewed with currently published data from light meson spectroscopy. It was previously shown in the CQ model that there existed several mass quanta m = 70 MeV, B = 140 MeV and X = 420 MeV, which were responsible for the quantization of meson *yrast* levels. The existence of a 70-MeV quantum was postulated by Mac Gregor and was shown to fit the Nambu empirical mass formula $m_n = (n/2)137m_e$, $n$ a positive integer. The 70-MeV quantum can be derived in three different ways: 1) pure electric coupling, 2) pure magnetic coupling, and 3) mixed electric and magnetic charges (dyons). Schwinger first introduced dyons in a magnetic model of matter. It is shown in this paper that recent data of new light mesons fit into the CQ model (a pure electric model) without the introduction of magnetic charges. However, by introducing electric and magnetic quarks (dyons) into the CQ model, new dynamical forces can be generated by the presence of magnetic fields internal to the quarks (dyons). The laws of angular momentum and of energy conservation are valid in the presence of magnetic charge. With the introduction of the Russell-Saunders coupling scheme into the CQ model, several new meson particles are predicted to exist. The existence of the $f_0(560)$ meson is predicted and is shown to fit current experimental data from the Particle Data Group listing. The existence of meson partners or groupings is shown.

PACS. 12.39.-x Phenomenological quark models - 12.40.Yx Hadron mass models and calculations - 14.80.Hv Magnetic monopoles

## 1 Introduction

In the course of researching particle physics over several decades, Mac Gregor [1] developed a comprehensive constituent-quark (CQ) model of elementary particles. It was shown that the CQ masses were directly related to the masses of the electron, muon, and pion. A connection was later discovered between the CQ masses of Mac Gregor's model and Nambu's empirical mass formula $m_n = (n/2)137m_e$, $n$ a positive integer and $m_e$ the mass of the electron [2]. Mac Gregor's esoteric notation included a 70 MeV quantum, a boson excitation B with the mass of the pion at 140 MeV, a fermion excitation F with a mass of 210 MeV or twice the muon mass, and a 420 MeV excitation quantum X. The 70 MeV quantum and the 420 MeV quantum X do not correspond to any observed particles but serve as the building blocks of mesons and baryons in the CQ model. These building blocks were <u>not</u> introduced in an *ad hoc* manner but were derived from experimental systematics and angular momentum systematics of hadron spectroscopy [3-4]. The mass regularities in meson and baryon spectra were identified independently of any theoretical viewpoint [5]. The mechanism for generating the CQ masses is discussed in Mac Gregor [1] and will be discussed further in Section 3. It is sufficient to say that we shall accept the evidence of Mac Gregor's CQ model from its previous agreement with earlier experimental data and that we shall present further evidence from the Particle Data Group listing [6]. The experimental evidence to be presented is limited to the meson spectrum only for the purpose of our study. The common mass-band structures of the baryons were also studied and can be found in Mac Gregor's work [1]. A sequel to the present paper on the subject of baryons can be found in Ref. [7].

In the CQ model, the mass of the resonance is determined mainly from the masses of the constituent quarks. Spin and orbital excitations can also contribute to the mass of the resonance with higher mass-states appearing at higher total angular momentum values. However, mesons with different J values appear in the same mass band in Fig. 2 of Mac

Gregor's work [1]. These meson resonances appear with accurate $J \sim M^2$ Regge trajectories for $J^P = 1^-$, $2^+$, and $3^-$. Likewise, the meson spectrum also exhibits non-Regge spacing with $J \sim M$ for the $2^+$, $3^-$, $4^+$, $5^-$ and $1^+$, $2^-$, $3^+$ yrast levels in Fig. 2 of Mac Gregor's paper. The spacing of the Regge-like and non-Regge levels is in accurate 420 MeV intervals, for which Mac Gregor assigned an excitation quantum $X^1 = 420$ MeV. The quantum X can appear with zero or non-zero units of angular momentum.

Additional sets of resonances are found to depend upon their quark content. For the CQ model, the constituent-quark basis states are calculated to be u(315), d(320), s(525), c(1575), and b(4725). The CQ model was postulated long before the advent of the top quark and has not been extended to the energy range for resonances involving the top quark. For purposes of our study, we limit our discussion to mesons below 2600 MeV. In addition to the excitation quantum X, there are mesons which exhibit equal spacing in the excitation quantum B = 140 MeV or the mass of the pion, as shown in Fig. 3 of Mac Gregor's work [1]. From the analysis in Ref. [1] and from previous work [3-5], Mac Gregor concluded that the excitation quantum B served as a fundamental mass unit.

In this paper, it is shown that there is exists a fine structure in one-half units of B or m = 70 MeV, separating meson band states. The mechanisms for generating this quantum m = 70 MeV will be discussed in Section 3. The evidence for this quantum is presented in Refs. [1, 3-5] and is shown in Table 1, which has the usual CQ model notation and which has the primary mesons $\pi^0$(134), $K^\pm$(494), and $\eta$(547) in the ground states. These mesons are listed with shell numbers (on the 70 MeV scale) in isospin space. Evidence of shell structure can be found in meson and baryon spectroscopy [8]. Earlier ideas of shell structure can be found in Ref. [9]. In Table 1, we list all the well-established mesons, underlining them. The not-so-well-established resonances are also included in the CQ model. This table is an update to Table VII in Mac Gregor's work [1] and has additional fine structure representing new states at the 1400 to 2450 MeV energy levels. These new states have m = 70 MeV regular spacings between them. This is a feature not explained in the Standard Model of elementary particle physics. It is noted that many if not all the known mesons below 2600 MeV fit into the CQ model. Moreover, in Table 2 we show a regrouping of the mesons from Table 1. On the 70-MeV scale, shell numbers are now identified in ordinary space in Table 2. In the paper by Mac Gregor [3], the 70-MeV excitation spacing can also be found for both mesons and baryons in his Fig. 8. Additional 70-MeV spacing is shown in Fig. 21 of Ref. [4]. The CQ notations for Tables 1 and 2 do <u>not</u> follow the rotational band rules established by Mac Gregor in Ref. [3], because we are only interested in the total CQ excitation masses for the energy levels shown in parentheses. Additional rules for hadronic binding energies and properties can be found in Ref. [4].

As shown in Tables 1 and 2, these mesons are grouped into excitation bands with different spins and isospins. For the sake of readers who may not be familiar with the notation of the CQ model, the energy bands are calculated from the total mass values as indicated by the symbols of the m, B, F and X quanta. For a notation such as $X_3$, this represents a molecular type writing by Mac Gregor and thus has value XXX = 420 + 420 + 420 = 1260 MeV, as shown in Table 1. The superscripts to these quanta represent spin J and parity P.

In studying Tables 1 and 2, we note that these mesons fit within the band structure of the CQ model, and yet there are spreads of particles within each band due to the estimates for peak masses and particle widths. A comparison of the CQ model and the particles in Table 1 could be made with the Standard Model. Comparisons to meson nonet membership and to meson fits to Regge theory are possible. However, these comparisons are not in the present scope of this paper and will be presented in a later paper. In this paper, we are concerned with the groupings of the particles in Table 1 and their spreads in energy. In the next section, we introduce quarks (dyons) with electric and magnetic charges into the CQ model, which allows for new dynamical forces in the meson spectrum. These mesons are studied in the CQ model, identifying their groupings or associated states.

## 2  CQ Model and Russell-Saunders Coupling

Mesons are composites of two quarks or quark-antiquark pairs and have been extensively studied as quarkonium and compared to nonrelativistic positronium-like bound states, which are observable as narrow resonances in electron-positron annihilation. If these quark-antiquark states are indeed comparable to atomic-like systems, then it would seem logical that the physics of the atomic scale would be applicable to the particle scale. However, the nature of the quarks would have dramatic effects as well on the physics. In the original CQ model, the quarks were of the Fermi-Yang type not the nucleon quarks of Gell-Mann and Zweig [3]. Other models of hadrons have been introduced with electric and magnetic quarks, involving spin J = 0 magnetic charges and spin J = ½ electric charges [10]. These are the quarks modeled in the present paper and are called dyons. The electric dipole moments (EDM) of these quarks are discussed in Ref. [10]. The idea that magnetic charge may be present in hadronic matter is attributed to Schwinger [11, 12]. The laws of the conversation of angular momentum and energy are valid in the presence of magnetic charge bound to electric charge in the nature of dyons, which are the quarks of this present paper. The presence of magnetic charges in elementary quarks allows for new dynamical forces to emanate from the magnetic fields generated by the sources. These internal fields to the quarks (dyons) may couple in the usual way of the weak Zeeman effect [13] or in the stronger sense of the Paschen-Back effect [14]. Thus, we are combining two models in the present paper: the CQ model (a pure electric coupling model) and the dyonium model of hadrons (a mixed electric and magnetic coupling model).

In studying the weak Zeeman or the strong Paschen-Back effect, we note that energy states may be split in the magnetic quantum number $m_j$ by the presence of *external* magnetic fields applied to a system of particles (or atoms). However, with magnetic charges present in the quarks, we have possible sources of *internal* magnetic fields, which can be applied to systems of particles (or mesons). In this section, we recall the Russell-Saunders coupling scheme [15] of two inequivalent particles and apply this scheme to the quarks in light meson spectroscopy. The Russell-Saunders coupling scheme assumes that the electrostatic interaction in atomic systems between two inequivalent electrons dominates the spin-orbit interaction. The orbital momenta and the spins of the particles couple separately to form $\boldsymbol{L} = \boldsymbol{L}_1 + \boldsymbol{L}_2$ and $\boldsymbol{S} = \boldsymbol{S}_1 + \boldsymbol{S}_2$. Then the total angular momentum is given by $\boldsymbol{J} = \boldsymbol{L} + \boldsymbol{S}$. For each $l$ and $s$, the $j$ values are $|l + s|$, ..., $|l - s|$. The combinations are shown in Table 3.

In Table 3, the number of states or number of $m_j$ values is shown. For the $^3P_{0, 1, 2}$ states, we have included the $(2j + 1)$ values or 1, 3, 5 for $j = 0, 1,$ and 2, respectively. If mesons are composites of two quarks (dyons), then we have two inequivalent quarks by reason of charge conjugation alone or by reason of flavor type in some cases.  Therefore, we would expect that the number of triplet *P*-states in meson systems to reflect the order of Table 3 with the $^3P_1$ states splitting into 3 separate levels and with the $^3P_2$ states splitting into 5 separate levels in the presence of a magnetic field.  We shall examine the experimental evidence for *P*-level splitting in light meson systems.

We utilize a scale of particle masses based upon the CQ model as found in Fig. 3 of the paper by Mac Gregor [1].  In fact, there are two distinct scales in the figure; one scale starts with the pion mass at 140 MeV and has steps of X = 420 MeV, and the other scale starts at zero and has steps of q = 315 MeV.  The X = 420 MeV scale has particle masses at $\pi(140)$, $\eta(547)$, $\eta'(958)$, $\eta(1440)$, $\eta(1760)$, and $\eta(2225)$.  The q = 315 MeV scale has particle masses at $\eta(1295)$, $\eta(1580)$, and D(1864); these particles can be identified at the mass levels 4q = 1260 MeV, 5q = 1575 MeV, and 6q = 1890 MeV, respectively.  These levels can be rearranged as follows: 4q = 3(420) = 630 + 630, a triple and transform reaction as noted in Ref. [1], so that the mass scales are interchangeable between mesons and baryons under the appropriate rules.  5q is more problematic, because this would seem to suggest that the meson $\eta(1580)$, a boson, is in fact a fermion from the quark content.  However, with the appropriate binding energy rules in Ref. [4], the predicted meson $\eta(1580)$ seems to fit better into the 1540 or 1610 MeV levels of Table 2.  Binding energies of the composite mesons are also discussed in Mac Gregor's work [1].  Finally, the D(1864) is easily identified by the rearrangement of the quark content as follows:  c = 5q = 1575 and u = q = 315; thus, D(1864) = c q, where c is the charm quark and q is the u or d antiquark.

The first set of particles corresponding to the 420 MeV scale is shown in Fig. 1. In Fig. 1, the experimental meson masses are indicated by solid lines and are taken from the Particle Data Group [6]. The vertical arrows represent an energy separation of about 420 MeV between states. We note a consistent pattern of energy separation between the spin-singlet and -triplet states. The lowest lying charmonium states are also shown for comparison, and a 420 MeV energy of separation is indicated by the arrows. There are distinct groupings as indicated by the arrows. The $\eta(2980)$ is associated with the $\chi_{c0}(3415)$ state from extensive study of the charmonium spectrum. Taking this pattern of energy separation to the lower meson states, we note that there are associated groupings or meson partners. $\eta(1295)$ is associated with $f_0(1710)$, $\eta'(958)$ with $f_0(1370)$, and $\eta(547)$ with $f_0(980)$. For the lowest lying state, $\pi(140)$ has a missing associated partner. A missing $f_0$ meson is shown at 560 MeV and is predicted to exist.

The $f_0(560)$ is a missing partner of the pion in the CQ model as inferred from Fig. 1. In the study of mesons in this mass range, there has been extensive debate in regards to the existence of the $\sigma(400\text{-}1200)$ meson. Numerous models have predicted the existence of the $\sigma$ meson, and in fact this meson is now identified as the $f_0(400\text{-}1200)$ scalar [16] and listed as the $f_0(600)$ by the Particle Data Group [6]. Van Beveren *et al* identified this scalar meson as the dynamically generated chiral partner of the pion [16]. It is interesting that the $f_0(560)$ is easily identified as a missing partner of the pion in Fig. 1. The predicted mass for the $f_0(560)$ can be compared to the experimental data with error bars. This comparison is shown in Fig. 2 as a function of publication date [6]. We note that the predicted mass of the $f_0(560)$ fits the experimental data.

The next set of particles corresponding to the 420 MeV scale is shown in Fig. 3. In Fig. 3, there is a consistent pattern of spin-spin and spin-orbit energy separation between the states. The set of particles η'(958), φ(1020), $f_0$(1370), $f_1$(1465), and $f_2$(1525) parallel the lowest lying charmonium states. This pattern is repeated with the set of particles η(1295), ω(1420), $f_0$(1710), $f_1$(1805), and $f_2$(1850). The low meson masses $f_1$(1465), $f_1$(1805) and $f_2$(1850) are predicted to exist. The pattern of energy separation is repeated with additional sets of particles as shown in Fig. 4. The solid lines represent the well-established particles taken from the Particle Data Group [6]. The dot-dashed lines represent mesons, which are predicted to exist from the symmetry of the pattern. For isospin I = ½ mesons, a similar pattern of energy separation is seen in Fig. 5, where a missing $D_0$ is predicted to exist with a mass of 2320 MeV. For all figures in this paper, the solid lines represent experimental meson masses, and the dashed-dot lines represent unobserved mesons, which are predicted to exist. In Figs. 1-5, the particles fit into the CQ model <u>without</u> the presence of magnetic charge, and the predicted mesons in these figures may exist independent of the ideas in the following text.

An additional set of mirror states to the charmonium states is shown in Fig. 6. These are isospin I = 0 mesons [17], and they have a symmetry about 2397 MeV as represented by the dashed line. The dashed line represents the mass of the classical Dirac magnetic monopole. The symmetry pattern has been attributed to the Zeeman splitting of meson states or more effectively the Paschen-Back effect [17]. We note that the spin-spin and spin-orbit energies are similar in separation between their respective states. In order to complete the symmetry in mass and isospin, a missing η meson is shown at about 1820 MeV. From the Particle Data Group listing [6], there exists a π(1800) which would fit into this bin at $J^{PC} = 0^{-+}$; however, the isospin I = 1 of the $\pi(1800)$ would break the isospin symmetry while filling in the mass level.

Starting with the first set of particles associated with η'(958) in Fig. 3, we search for particle states, which satisfy the number of states for Russell-Saunders coupling in Table 3. The results of this search are shown in Fig. 7. In Fig. 7, there are indications of particle states, which fit the Russell-Saunders coupling scheme with 1, 3, and 5 states for the triplet *P*-states ($l = 1$, $s = 1$). With the introduction of $a_2(1320)$ into these states, there is isospin-symmetry breaking, and there is then evidence for 5 mass splittings in the $^3P_2$ states consistent with the Russell-Saunders coupling scheme. Experimental masses are indicated with solid lines. Those indicated by dashed-dot lines are predicted to exist.

We can repeat the process of searching for particle states, which fit the Russell-Saunders coupling scheme by utilizing the scale determined from the CQ model. These results are shown in Figures 8 to 15. A total of 9 figures, including Fig. 7, represent evidence consistent with the Russell-Saunders coupling scheme of Table 3. In Fig. 8, we note that for $f_0(1500)$ there does not appear to be associated mesons in the $0^{-+}$ and $1^{--}$ bins. In Fig. 9, we introduce the $a_1(1640)$ to fill the mass level and isospin-symmetry is broken. We note the consistent pattern of states, which fit the Russell-Saunders coupling scheme in Fig. 9. Similarly, we introduce the $a_2(1750)$ into the pattern for Fig. 10, which has mesons associated with the η(1440) meson. For Figures 11 to 13, the same pattern of states are found grouped with an associated meson. There is no indication of isospin-symmetry breaking to fill the particle states for Figures 11 to 13. In Fig. 11, we indicated the space for X(1900) in the $J^{PC} = 2^{++}$ bin. There is some evidence for a $f_2(1910)$ meson as noted in the Particle Data Group listing [6]. In Fig. 14, there is no hint of isospin-symmetry breaking for I = ½ mesons.  In Fig. 15, there is indication of isospin-symmetry breaking with the η(2225) filling the $0^{-+}$ bin.

## 3  Physics Beyond the Standard Model

It was noted in the last section that the evidence for Russell-Saunders or *LS* coupling is derived from the light meson spectra of Figures 7 to 15.  If these spectra can be shown to satisfy the Lande interval rule, which is widely used in atomic, molecular and nuclear physics, then we have evidence for the presence of Russell-Saunders or *LS* coupling.  The Lande interval rule is the test.  For the *P*-states ($J = 2$), the Lande interval rule predicts a mass splitting or ratio of 2.0 for the states within the same multiplet.  As shown in Table 4, we have calculated a total of 10 different mass splittings for comparison to the theoretical Lande ratio of 2.0 in these states.  The Lande interval rule is satisfied even though we have 4 out of 10 states in Table 4 with missing mesons to be filled into the scheme.  The values in Table 4 are comparable to those found in atomic systems [18].  A more extensive study of the Lande ratio for the P-states would involve a complete derivation of the Lande factor g in the Zeeman energy splitting of these meson states:

$$\Delta E = - \mu_B \, g \, M_J \, B. \tag{1}$$

The evidence of Russell-Saunders coupling is suggestive in particle physics.  The existence of Russell-Saunders coupling, which is lifted by the presence of a magnetic field in Figures 6 to 15, is evidence for strong internal magnetic fields within the composite mesons and represents physics beyond the Standard Model.  The effects of these strong internal magnetic fields are shown in Figures 6 to 15, and the sources of these internal magnetic fields B is attributed to the magnetic charges within the quarks (dyons) in the present model.

The CQ model is a pure electric coupling model; namely, it encompasses only quarks with electric charges. If the Russell-Saunders coupling scheme were introduced into the CQ model <u>without</u> the presence of magnetic charge, then the internal angular momentum of the mesons would <u>not</u> be conserved. A magnetic field in extra-dimensional space [19] would need to be introduced as an explanation for these effects and in order to conserve angular momentum. However, this would lead us into physics beyond the Standard Model.

In the CQ model, the generation of particle masses comes from the creation of a spinless boson m = 70 MeV. This boson occurs in two isotopic-spin states [1]. The mass is derived by the formula:

$$m = m_e/\alpha = 70 \text{ MeV}. \qquad (2)$$

Eq. (2) represents a pure electric coupling with the fine structure constant $\alpha = 1/137$. This is the energy scale first noted by Nambu [20] and developed later by Mac Gregor [3-4], who first introduced a relativistic spinning sphere model for the 70 MeV quantum (see reference 123 in our Ref. [3]). With many problems in physics, an electromagnetic duality can exist between electric and magnetic charges with the appropriate transformation laws. Indeed, the 70 MeV quantum can also be derived from the classical Dirac magnetic monopole mass and the strong coupling constant $\alpha_s$ between magnetic charges. Sawada has attributed the strength of p-p scattering at low energies to the present of magnetic charge within hadrons [21-24]. The strong coupling constant is [24]:

$$\alpha_s = 34.25 \qquad (3)$$

for a composite particle with magnetic charges of opposite signs. The classical Dirac magnetic monopole mass is m = 2397 MeV, which is derived from conservation of energy density. Electromagnetic duality suggests the replacement of the electron mass in Eq.(2) with the mass of the classical Dirac magnetic monopole m = 2397 MeV and replacement of the fine structure constant $\alpha$ = 1/137 with the strong coupling constant $\alpha_s$ = 34.25 in Eq. (3):

$$m = m_{\text{Dirac monopole}}/\alpha_s = 70 \text{ MeV}. \tag{4}$$

As expected, the 70 MeV quantum can be derived from electromagnetic duality with the appropriate coupling constant. In fact, the 70 MeV quantum can also be derived by a third method with mixed electric and magnetic charges [2]. In Ref. [2], the author derived Nambu's empirical mass formula which is inclusive of the 70 MeV quantum (see Eq. (31) and Table I in Ref.[2]).

## 4 Conclusion

In this paper, we reviewed the constituent-quark model of Mac Gregor [1], derived two particle mass scales to identify meson spectra, and presented experimental data derived from the Particle Data Group [6]. The particle mass scales were derived from Nambu's empirical mass formula [2] and were noted long before our derivation by Mac Gregor [1]. The 70 MeV quantum and the 420 MeV quantum served as the building blocks of mesons and baryons in the CQ model. These building blocks were not introduced in an *ad hoc* manner but were derived from experimental systematics and angular momentum systematics of hadron spectroscopy [3-4]. We presented experimental evidence for mesons, which fit into the energy scales of the constituent-quark model (a pure electric coupling model). We introduced magnetic charges into the quarks by combining the CQ model with the dyonium model of hadrons by Schwinger. The problem of electric dipole moments (EDM) in such models has been discussed elsewhere [10, 17]. The presence of magnetic charge inside the quarks allows for internal magnetic fields, which admit new possible physics in elementary particles. These magnetic fields allow for the construction of Russell-Saunders splitting of meson states, which has never been seen before now in particle physics. This evidence was tested by the Lande interval rule. The tabulated data satisfied the Lande interval rule. It was suggested that strong internal magnetic fields of the composite mesons are responsible for the mass splittings of Figures 7 to 15. These internal magnetic fields produce a Zeeman or Paschen-Bach effect that lifts the degeneracy of the *P*-states and generates distinct particles. Thus, we are left with light meson spectra which fit the Russell-Saunders coupling scheme and which satisfy the Lande interval rule.

***Note Added in Proof.***

Evidence for the presence of magnetic charge or monopoles is being currently studied by experimentalists [25]. A current bibliography on magnetic monoples may be found in [26]. Since the writing of this paper, there have been a number of discoveries by various groups. First, the BES Collaboration has discovered the possible existence of a pseudoscalar state with a mass of 1859 MeV [27]. This is consistent with the prediction of an eta (1820) meson as shown in Fig. 6. Second, the Belle Collaboration has reported the discovery of a broad scalar $D_0^*(2308)$ in a study of $B^-$ decays [28]. This is consistent with the prediction of a $D_0(2320)$ in Fig. 5. Third, the LEPS Collaboration [29] has discovered a S = +1 baryon resonance at 1540 MeV, which has been confirmed by the CLAS Collaboration [30]. In Section 2, we noted that the 5q = 1575 MeV state would be problematic for the $\eta(1580)$ meson and that this meson would seem to fit better into the 1540 or 1610 MeV levels of Table 2. In the CQ model, the nucleon quark mass is q =

315 MeV. With consideration of 2-3% binding energies, the pentaquark state would be a little less than 5q = 1575 MeV and would be consistent with the discovery of the S = +1 baryon resonance at 1540 MeV. Mac Gregor noted long ago that both mesons and baryons share the same constituent quark mass bands. Finally, the author discovered in a literature search a quantum-mechanical derivation of the Zeeman effect for an electric charge − magnetic monopole system by Barker and Granziani [31]. This derivation is shown in their Eqs. (35S) and (36M) in Ref. [31]. Thus, the idea of Zeeman splitting is possible in hadron spectroscopy in the presence of magnetic charge.

As a final note, we would point out the great success of QCD and the Standard Model of particle physics. Attempts to construct the Standard Model with magnetic monopoles have been ongoing by Vachaspati [32] and in the Dual Standard Model by Liu *et al.* [33].

## Acknowledgement

The author wishes to thank Dr. Malcolm Mac Gregor of the University of California's Lawrence Livermore National Laboratory for his encouragement to pursue the CQ Model, and he wishes to thank Dr. Paolo Palazzi of CERN for his interest in the work and for e-mail correspondence.

Table 1. Meson and kaon resonances for all the well-established resonances are underlined [6]. The not-so-well-established resonances are included. Shell numbers are identified in isospin space as 2, 7, and 8 times the mass m = 70 MeV.

| Shell Numbers: (on 70 MeV scale) | | 2m | 7m | 8m |
|---|---|---|---|---|
| Isospin | | I = 1 | I = ½ | I = 0 |
| Ground states $J^{PC}$ (in MeV) | | $\underline{\pi^0(134.98)}0^{-+}$ $\underline{\pi^\pm(139.5)}0^-$ | $\underline{K^\pm(494)}0^-$ $\underline{K^0(498)}0^-$ | $\underline{\eta(547)}0^{-+}$ $f_0(560)0^{++}$ |

| CQ excitation band (in MeV) | Level | | | |
|---|---|---|---|---|
| F = 3m | (210) | | | $\underline{\omega(782)}1^{--}$ |
| FB = 5m | (350) | | $\underline{K^*(892)}1^-$ | |
| X = **6m** | **(420)** | | | $\underline{\boldsymbol{\eta'(958)}}0^{-+}$, $\underline{f_0(980)}0^{++}$ |
| FBB = 7m | (490) | | | $\underline{\phi(1020)}1^{--}$ |
| XF = 9m | (630) | $\underline{\rho(770)}1^-$ | | $\underline{h_1(1170)}1^{+-}$ |
| XBB = 10m | (700) | | | $\underline{f_2(1270)}$, $\underline{f_1(1285)}$, $\underline{\eta(1295)}$ |
| XFB = 11m | (770) | | $\underline{K_1(1270)}1^+$ | |
| XX = **12m** | **(840)** | $\underline{a_0(985)}0^{++}$ | | $\underline{f_0(1370)}$, $\underline{\omega(1420)}$, $\underline{f_1(1420)}$, $f_2(1430)$, $\underline{\boldsymbol{\eta(1440)}}$ |
| XXm = 13m | (910) | | $\underline{K_1(1400)}$, $\underline{K^*(1410)}$ $\underline{K_0^*(1430)}$, $\underline{K_2^*(1430)}$ | $\underline{f_0(1500)}$ |
| XXB = 14m | (980) | | K(1460) | $f_1(1510)$, $\underline{f_2'(1525)}$, $f_2(1565)$ |
| XXBm = 15m | (1050) | | $K_2(1580)$ | $h_1(1595)$, $f_2(1640)$, $\eta_2(1645)$, $\underline{\omega(1650)}$ |
| XXBB = 16m | (1120) | $\underline{b_1(1235)}$, $\underline{a_1(1260)}$ | K(1630), $K_1(1650)$ | $\underline{\omega_3(1670)}$, $\underline{\phi(1680)}$, $\underline{f_0(1710)}$ |
| XXFB = 17m | (1190) | $\underline{\pi(1300)}$, $\underline{a_2(1320)}$ | $\underline{K^*(1680)}$ | $\eta(1760)$ |
| XXX = **18m** | **(1260)** | $h_1(1380)$, $\pi_1(1400)$ | $\underline{K_2(1770)}$, $\underline{K_3^*(1780)}$ | $f_2(1810)$, $\underline{\phi_3(1850)}$ |
| $X_3m = 19m$ | (1330) | $a_0(1450)$, $\rho(1450)$ | $K_2(1820)$, $K(1830)$ | $\eta_2(1870)$, $f_2(1910)$ |
| $X_3B = 20m$ | (1400) | | | $f_2(1950)$ |
| $X_3F = 21m$ | (1470) | $\pi_1(1600)$, $a_1(1640)$ $K_0^*(1950)$, $K_2^*(1980)$ | | $f_2(2010)$, $f_0(2020)$, $f_3(2050)$ |
| $X_3Fm = 22m$ | (1540) | $\pi_2(1670)$, $\rho_3(1690)$ $\rho(1700)$, $a_2(1700)$ | $K_4^*(2045)$ | $f_0(2100)$ |
| $X_3FB = 23m$ | (1610) | | | $f_2(2150)$, $f_0(2200)$ |
| $X_4 = \mathbf{24}m$ | (**1680**) | $\pi(1800)$ | | $f_J(2220)$, $\boldsymbol{\eta(2225)}$ |
| $X_4m = 25m$ | (1750) | $\rho(1900)$ | $K_2(2250)$ | $f_2(2300)$, $f_4(2300)$, $f_0(2330)$, $f_2(2340)$ |
| $X_4B = 26m$ | (1820) | $\rho_3(1990)$ | $K_3(2320)$ | |
| $X_4F = 27m$ | (1890) | $a_4(2040)$ | $K_5^*(2380)$ | |
| $X_4Fm = 28m$ | (1960) | $\pi_2(2100)$ | | $f_6(2510)$ |
| $X_4FB = 29m$ | (2030) | $\rho(2150)$ | $K_4(2500)$ | |
| $X_5 = \mathbf{30}m$ | (**2100**) | $\rho_3(2250)$ | | |
| $X_5m = 31m$ | (2170) | $\rho(2350)$ | | |
| $X_5F = 33m$ | (2310) | $a_6(2450)$ | | |

Table 2. Meson and kaon resonances for all the well-established resonances are underlined. The not-so-well-established resonances are included. Resonances indicated by boldface are new particles, which were not listed by Mac Gregor in 1990 [1]. Shell numbers are identified in ordinary space. η(1580) is predicted to exist.

| Isospin | | I = 1 | I = ½ | I = 0 |
|---|---|---|---|---|
| CQ excitation band | Level | | | |
| Shell Numbers (on m = 70 MeV scale) | | | | |
| B = 2m | (140) | $\pi^0$(134.98)0$^{-+}$  $\pi^{\pm}$(139.5)0$^{-}$ | | |
| Xm = 7m | (490) | | $K^{\pm}$(494)0$^{-}$  $K^0$(498)0$^{-}$ | |
| XB = **8**m | (**560**) | | | $\eta$(547)0$^{-+}$  $f_0$(560)0$^{++}$ |
| XFB = 11m | (770) | $\rho$(770)1$^{--}$ | | $\omega$(782)1$^{--}$ |
| XX = 12m | (840) | | K*(892)1$^{-}$ | |
| XXB = **14**m | (**980**) | $a_0$(985)0$^{++}$ | | **$\eta'$(958)**0$^{-+}$, $f_0$(980)0$^{++}$ |
| XXF = 15m | (1050) | | | $\phi$(1020)1$^{--}$ |
| XXFB = 17m | (1190) | | | $h_1$(1170)1$^{+-}$ |
| XXX = 18m | (1260) | $b_1$(1235), $a_1$(1260) | $K_1$(1270) | $f_2$(1270), $f_1$(1285)  $\eta$(1295) |
| X$_3$m = 19m | (1330) | $\pi$(1300), $a_2$(1320) | | |
| XXFFB = **20**m | (**1400**) | $h_1$(1380), $\pi_1$(1400) | $K_1$(1400), K*(1410)  $K_0$*(1430), $K_2$*(1430) | $f_0$(1370), $f_1$(1420)  $\omega_1$(**1420**), $f_2$(1430)  **$\eta$(1440)** |
| X$_3$F = 21m | (1470) | **$a_0$(1450)**, **$\rho_1$(1450)** | **$K_0$(1460)** | **$f_0$(1500)** |
| XXFFBB = 22m | (1540) | | **$K_2$(1580)** | $f_1$(1510), **$f'_2$(1525)**,  **$f_2$(1565)**, $\eta$(1580) |
| XXXFB = 23m | (1610) | **$\pi_1$(1600)**, **$a_1$(1640)** | **K(1630)**, **$K_1$(1650)** | **$h_1$(1595)**, **$f_2$(1640)**,  **$\eta_2$(1645)**, **$\omega$(1650)** |
| $X_3^{2+,1+} X^{1-} = 24m$ | (1680) | $\pi_2(1670)$, $\rho_3(1690)$ $\rho(1700)$, **$a_2(1700)$** | **K\*(1680)** | $\omega_3(1670)$, $\phi(1680)$, **$f_0(1710)$** |
| $mX_3^{2+,1+} X^{1-} = 25m$ | (1750) | | **$K_2(1770)$**, **$K_3^*(1780)$** | $\eta(1760)$ |
| XXXFFB = **26**m | (**1820**) | **$\pi(1800)$** | **$K_2(1820)$**, **$K(1830)$** | **$f_2(1810)$**, **$\phi_3(1850)$** |
| XXXFFF 27m | (1890) | **$\rho(1900)$** | | **$\eta_2(1870)$, $f_2(1910)$** |
| XXXFFBB = 28m | (1960) | **$\rho_3(1990)$** | **$K_0^*(1950)$, $K_2^*(1980)$** | **$f_2(1950)$** |
| XXXFFFB = 29m | (2030) | **$a_4(2040)$** | **$K_4^*(2045)$** | **$f_3(2010)$**, **$f_0(2020)$**, **$f_4(2050)$** |
| $X_3^{2+,1+} X^{1-} X^{1-} = 30m$ | (2100) | **$\pi_2(2100)$** | | **$f_0(2100)$** |
| XXXXFFm = 31m | (2170) | **$\rho(2150)$** | | **$f_2(2150)$, $f_0(2200)$** |
| XXXXFFB = **32**m | (**2240**) | **$\rho_3(2250)$** | **$K_2(2250)$** | **$f_J(2220)$**, **$\eta(2225)$**, |
| XXXXXF = 33m | (2310) | **$\rho(2350)$** | **$K_3(2320)$** | **$f_2(2300)$**, **$f_4(2300)$**, **$f_0(2330)$**, **$f_2(2340)$**, |
| XXXXXFm = 34m | (2380) | | **$K_5^*(2380)$** | |
| XXXXXFB = 35m | (2450) | **$a_6(2450)$** | | |
| $X_6^{1-} = 36m$ | (2520) | | **$K_4(2500)$** | **$f_6(2510)$** |

Table 3.  Russell-Saunders Coupling of Two Inequivalent Particles.

| $l$ | $s$ | $j$ | Spectral Terms | Number of States in a Magnetic Field (Number of $m_j$ Values) |
|---|---|---|---|---|
| 2 | 1 | 3, 2, 1 | $^3D_{1,2,3}$ | $3 + 5 + 7 = 15$ |
| 2 | 0 | 2 | $^1D_2$ | 5 |
| 1 | 1 | 2, 1, 0 | $^3P_{0,1,2}$ | $1 + 3 + 5 = 9$ |
| 1 | 0 | 1 | $^1P_1$ | 3 |
| 0 | 1 | 1 | $^3S_1$ | 3 |
| 0 | 0 | 0 | $^1S_0$ | 1 |
| | | | ---------- | ------------ |
| | | | 10 States | 36 States |

Table 4.  Calculations of the mass splittings for *P*-states ($J = 2$) associated with the mesons η'(958), η(1295), η(1440), f$_0$(1500), and η(1580).

| η'(958) | Upper states:  [f$_2$(1525) − f$_2$(1380)]/[f$_2$(1430) − f$_2$(1380)] = 2.9 |
|---|---|
| | Lower states: [f$_2$(1380) − f$_2$(1270)]/[f$_2$(1380) − a$_2$(1320)] = 1.8 |
| η(1295) | Upper states:  [X(1850) − a$_2$(1750)]/[f$_2$(1810) − a$_2$(1750)] = 1.7 |
| | Lower states: [a$_2$(1750) − f$_2$(1565)]/[a$_2$(1750) − f$_2$(1640)] = 1.7 |
| η(1440) | Upper states:  [f$_2$(2010) − X(1900)]/[f$_2$(1950) − X(1900)] = 2.2 |
| | Lower states: [X(1900) − a$_2$(1750)]/[X(1900) − f$_2$(1810)] = 1.7 |
| f$_0$(1500) | Upper states:  [f$_2$(1640) − f$_2$(1525)]/[f$_2$(1565) − f$_2$(1525)] = 2.9 |
| | Lower states: [f$_2$(1525) − f$_2$(1380)]/[f$_2$(1525) − f$_2$(1430)] = 1.5 |
| η(1580) | Upper states:  [f$_2$(2150) − f$_2$(2010)]/[X(2070) − f$_2$(2010)] = 2.3 |
| | Lower states: [f$_2$(2010) − X(1900)]/[f$_2$(2010) − f$_2$(1950)] = 1.8 |

## Figure Captions

**Fig. 1.** Experimental meson masses, indicated as solid lines, are taken from the Particle Data Group [6]. The vertical arrows indicate a separation energy of about 420 MeV. The lowest lying charmonium states are shown. Note the consistent pattern of the separation energy between the spin-singlet and -triplet states. The $f_0(560)$ is predicted to exist.

**Fig. 2.** Experimental masses, taken from the Particle Data Group, are plotted as a function of publication date and are indicated as solid points with error bars. The $f_0(560)$ is indicated by the dashed line.

**Fig. 3.** Low meson masses are shown for comparison to the lowest lying charmonium states. Note the consistent pattern of spin-spin and spin-orbit energy separations between the states. The low meson masses $f_1(1465)$ and $f_1(1805)$ are predicted to exist. The $f_2(1850)$ meson, as indicated by the doted line, has some evidence for its existence.

**Fig. 4.** Another set of low meson masses is shown in comparison to the charmonium states. Note the consistent pattern of energy separations between the groups. The X(1880) state is predicted to exist.

**Fig. 5.** Another set of low meson masses is shown in comparison to the lowest lying charmonium states. Note the consistent pattern of energy separations between the groups. The $D_0(2320)$ is predicted to exist.

**Fig. 6.** Zeeman splitting of isospin $I = 0$ mesons. The dashed line represents the mass of the Dirac monopole or 2397 MeV from classical calculations. The experimental masses are indicated by solid lines. The η meson at 1820 MeV is predicted to exist. Note the symmetry of the low mass mesons with the charmonium states about the dashed line.

**Fig. 7.** The set of low meson masses connected with the η'(958) state. Experimental meson masses are indicated with solid lines. Those indicated by dashed-dot lines are predicted to exist and fit the Russell-Saunders coupling scheme.

**Fig. 8.** The set of low meson masses without connection to a singlet state. Experimental meson masses are indicated with solid lines. Those indicated by dashed-dot lines are predicated to exist and also fit the Russell-Saunders coupling scheme.

**Fig. 9.** The set of low meson masses connected with the η(1295) state. Solid lines represent experimental meson masses taken from the Particle Data Group [6]. Dashed-dot lines are mesons which are predicted to exist. Note the pattern of mass splittings which fit the Russell-Saunders coupling scheme.

**Fig. 10.** The set of low meson masses associated with the η(1440) state. Solid lines indicate experimental masses taken from the Particle Data Group [6]. Dashed-dot lines represent mesons which are predicted to exist.

**Fig. 11.** The set of low meson masses connected with the η(1580) state, which is predicted to exist.  Experimental meson masses are indicated with solid lines.  Those indicated by dashed-dot lines are predicted to exist.  The $^3P_2$ states fit the Russell-Saunders coupling scheme.

**Fig. 12.** The set of low meson masses associated with the $f_0$(2100) state.  The notation is the same as in Fig. 10.

**Fig. 13.** The set of low meson masses associated with the η(1760) state.  The notation is the same as in Fig. 10.

**Fig. 14.** The set of low meson masses connected with the D(1864) state.  The notation is the same as in Fig. 10.

**Fig. 15.** The set of low meson masses associated with the η(2225) state.  The notation is the same as in Fig. 10.

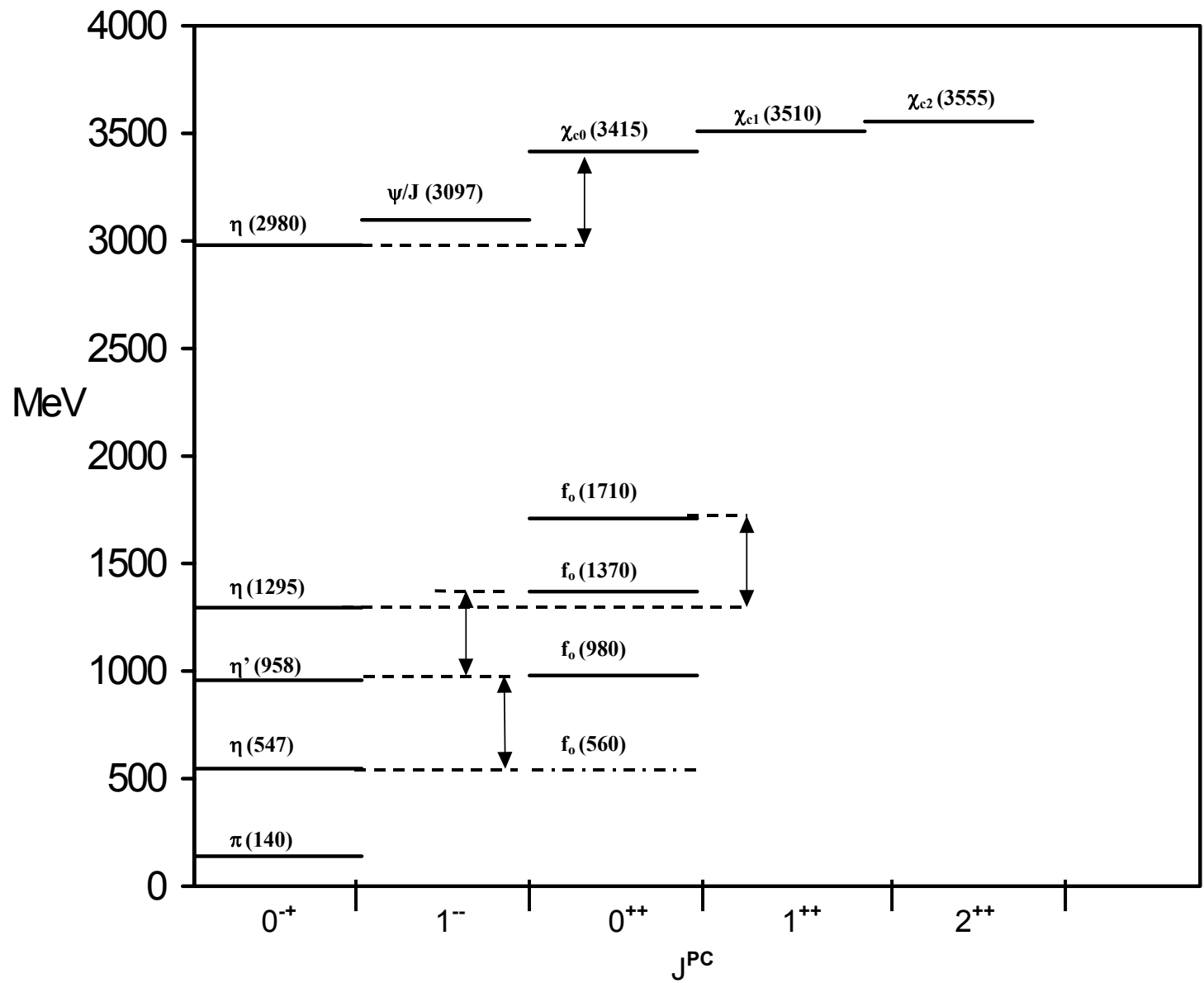

Fig. 1.

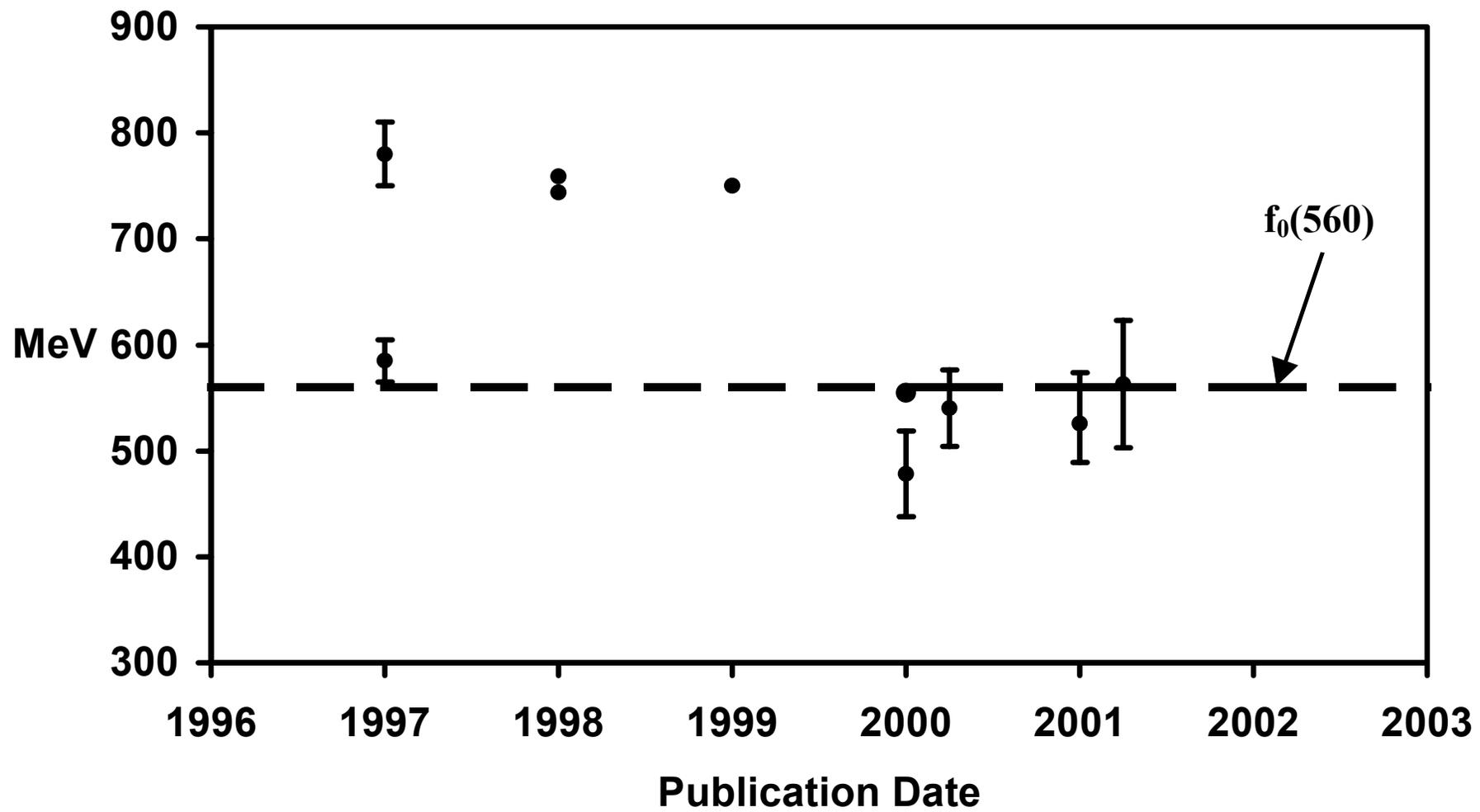

**Fig. 2.**

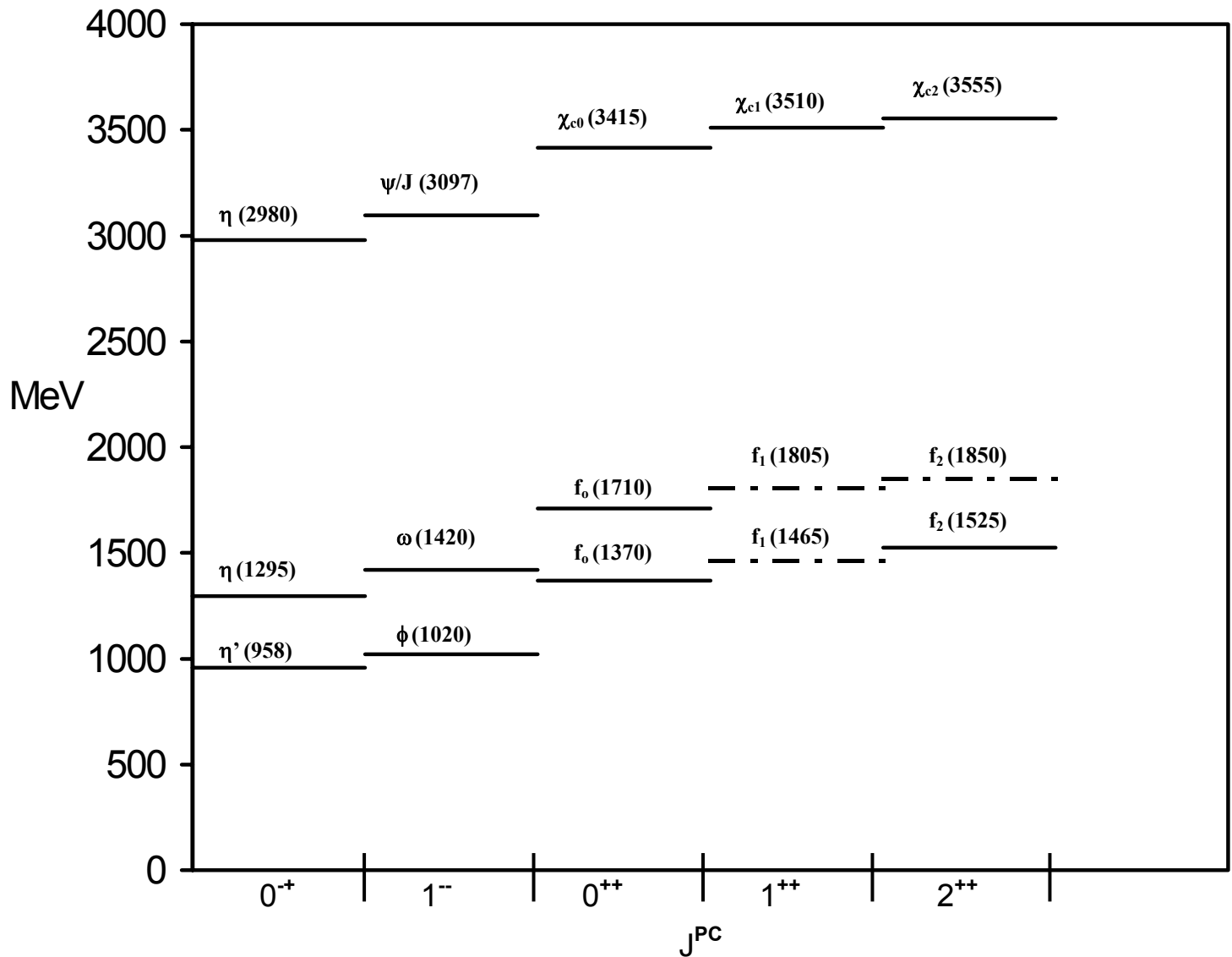

**Fig. 3.**

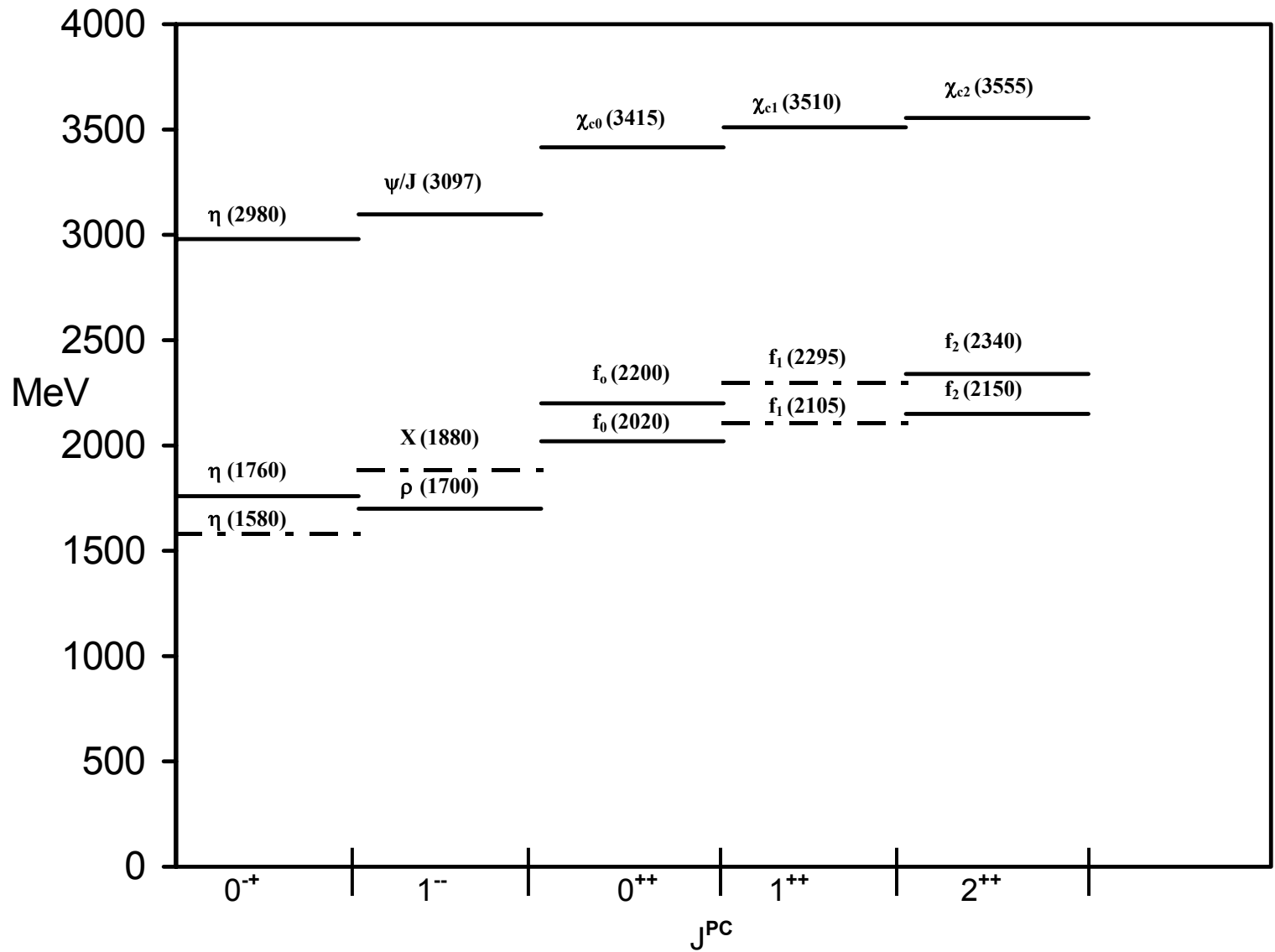

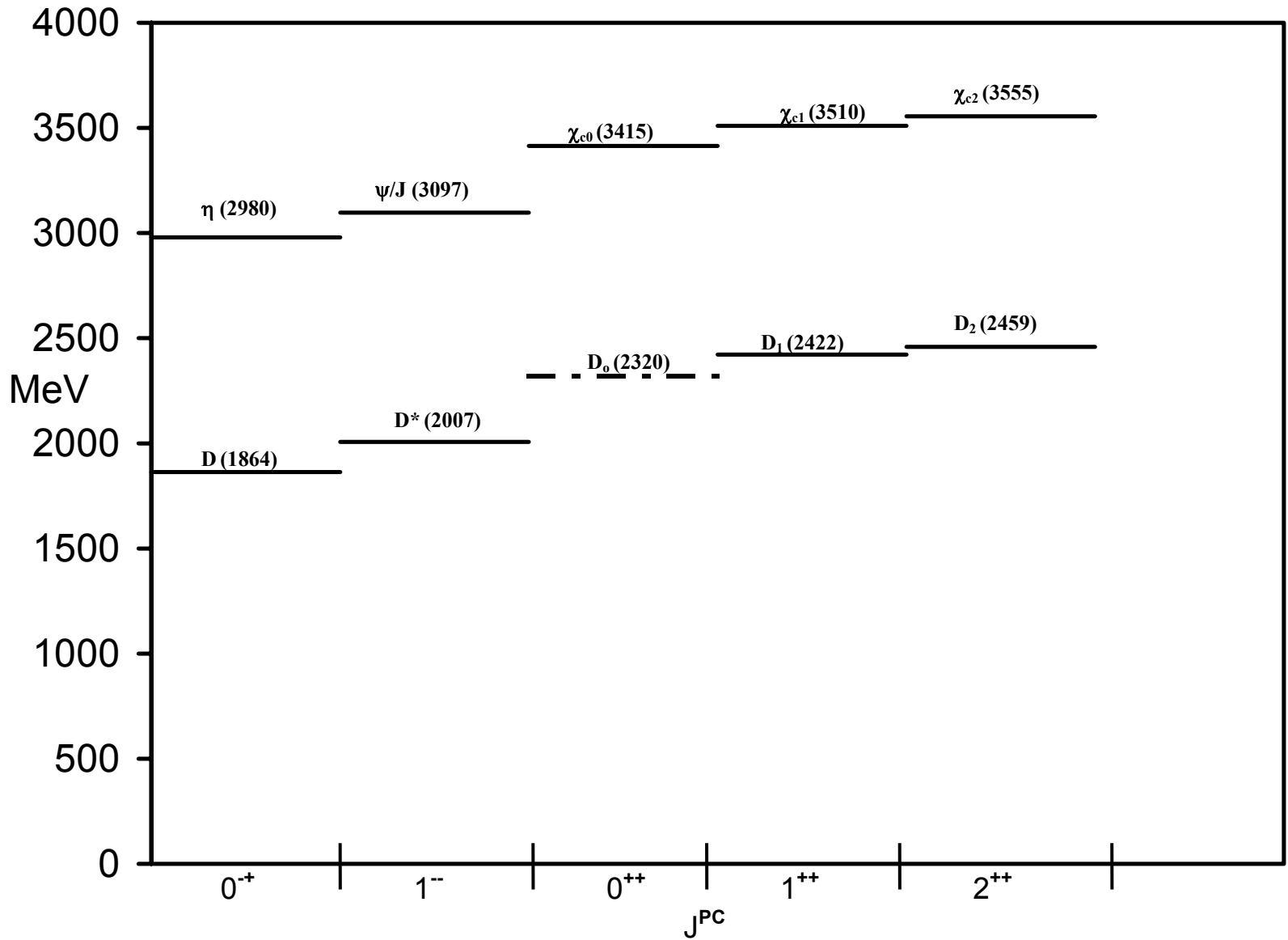

Fig. 5.

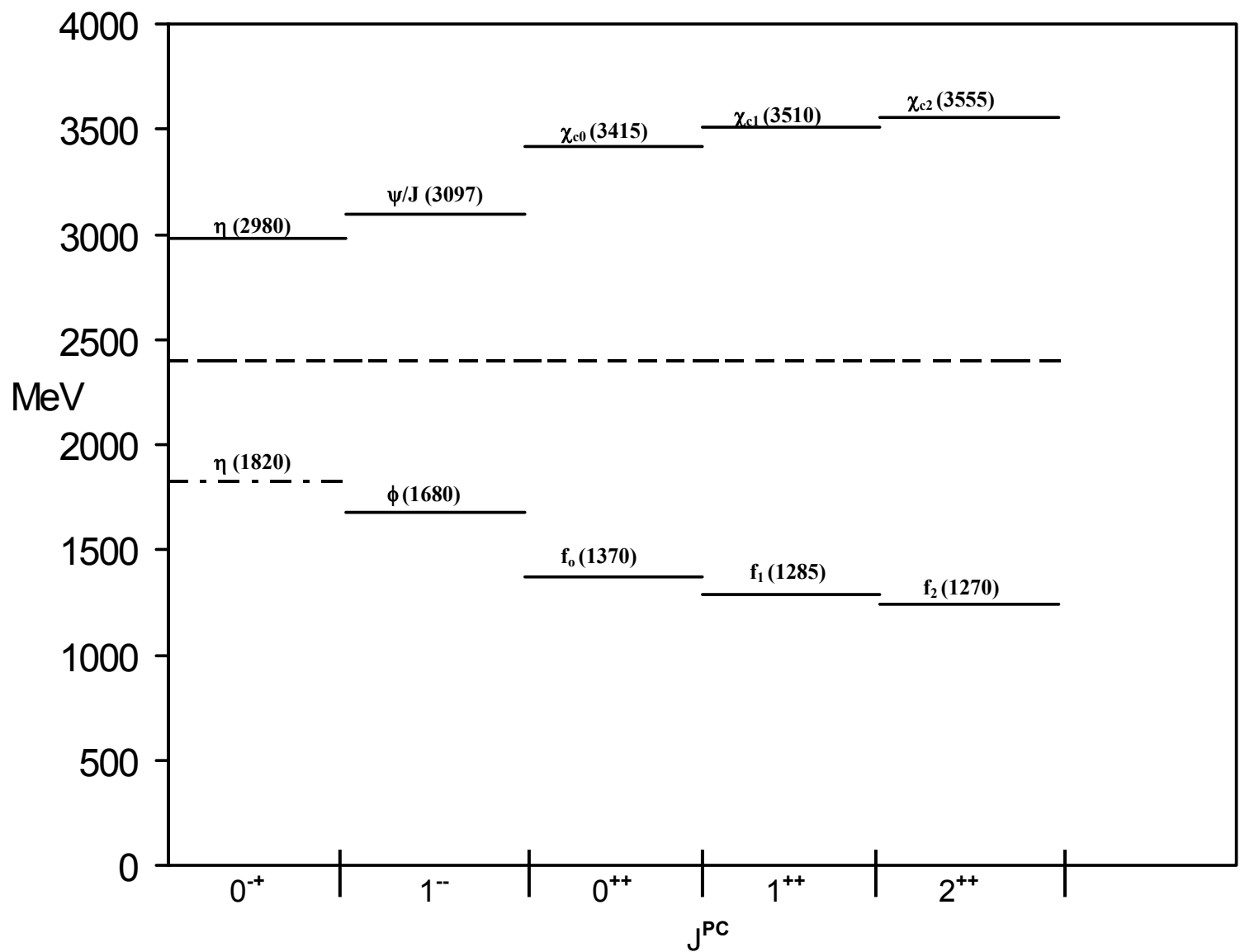

Fig. 6.

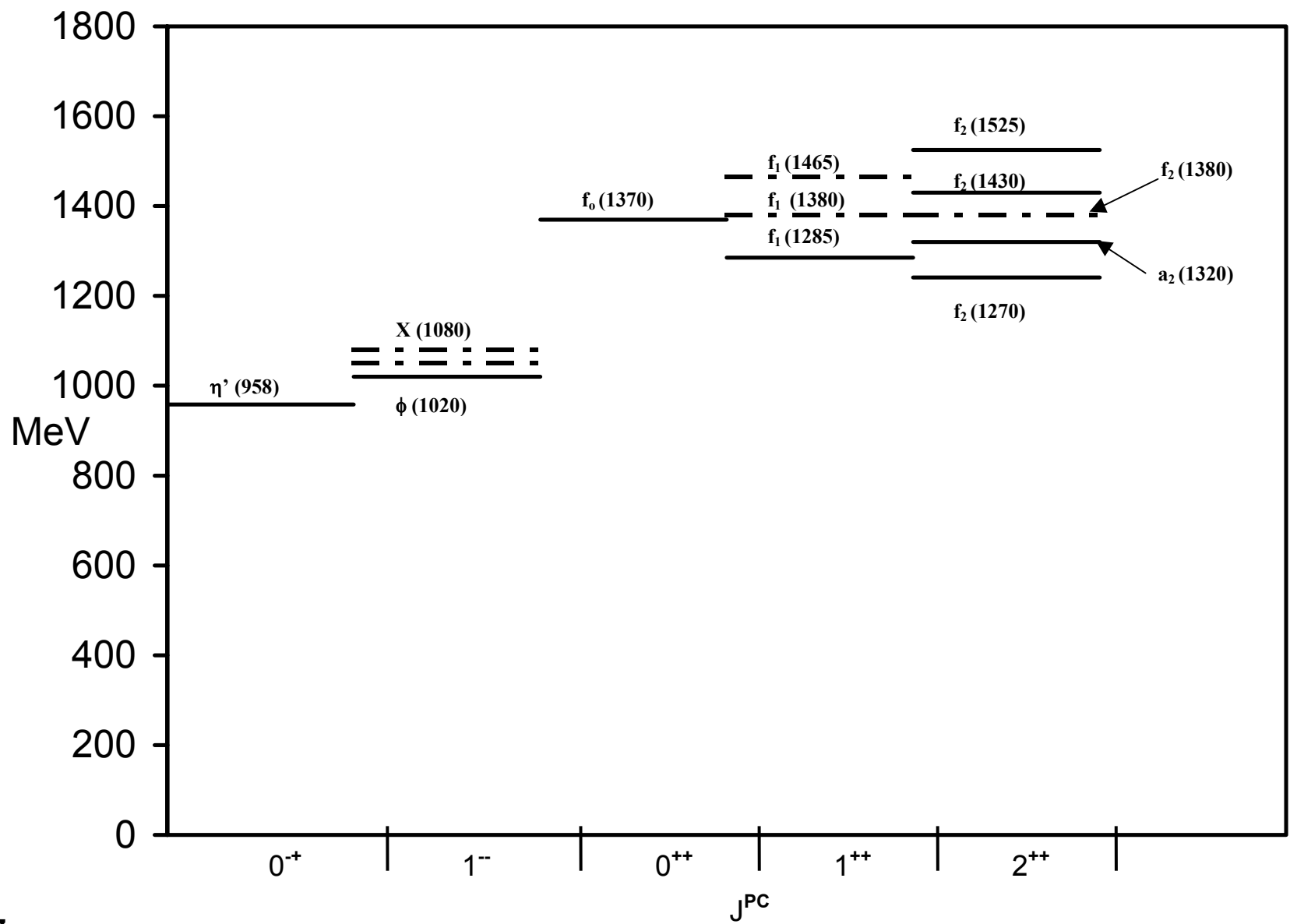

**Fig. 7.**

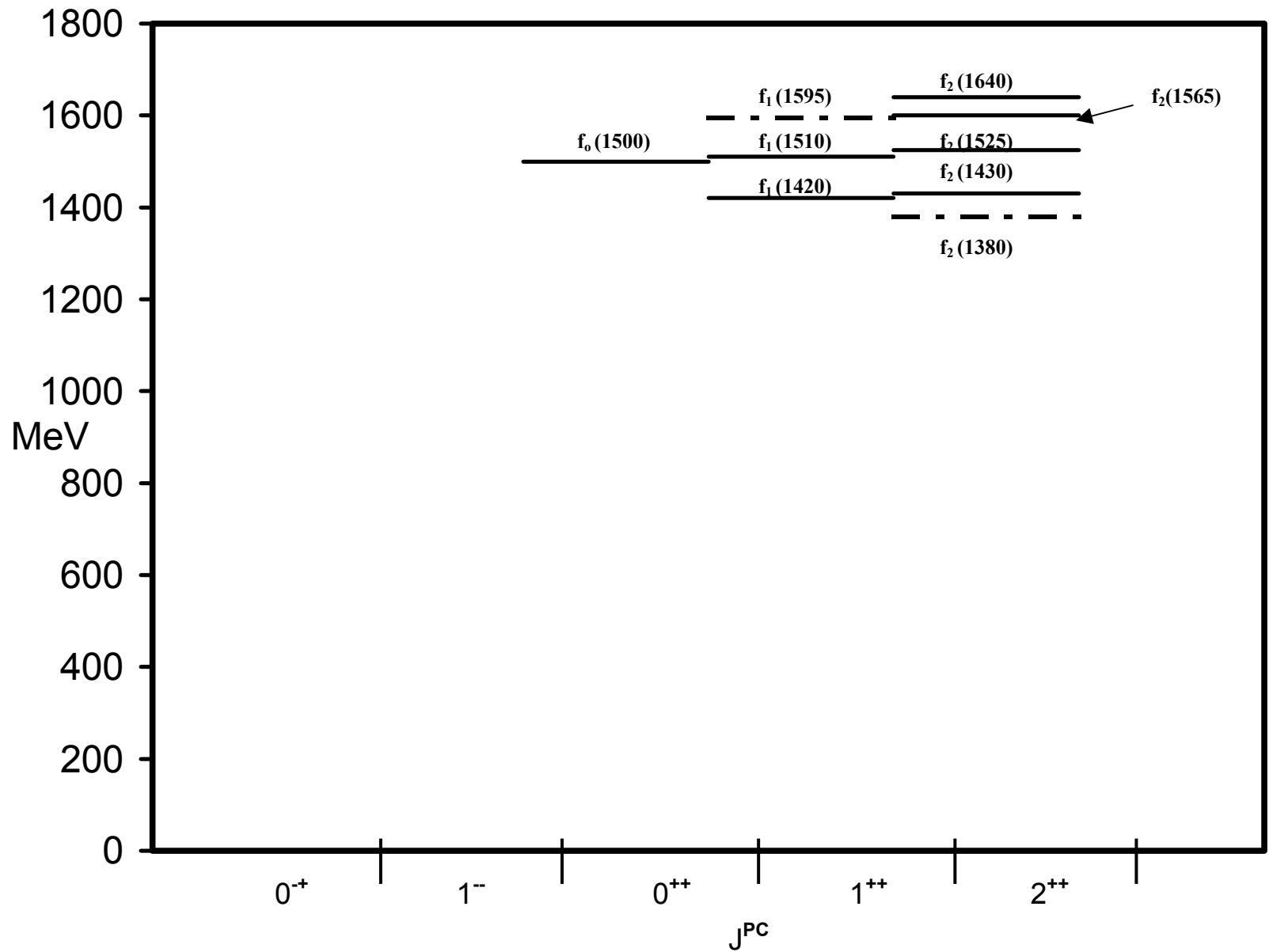

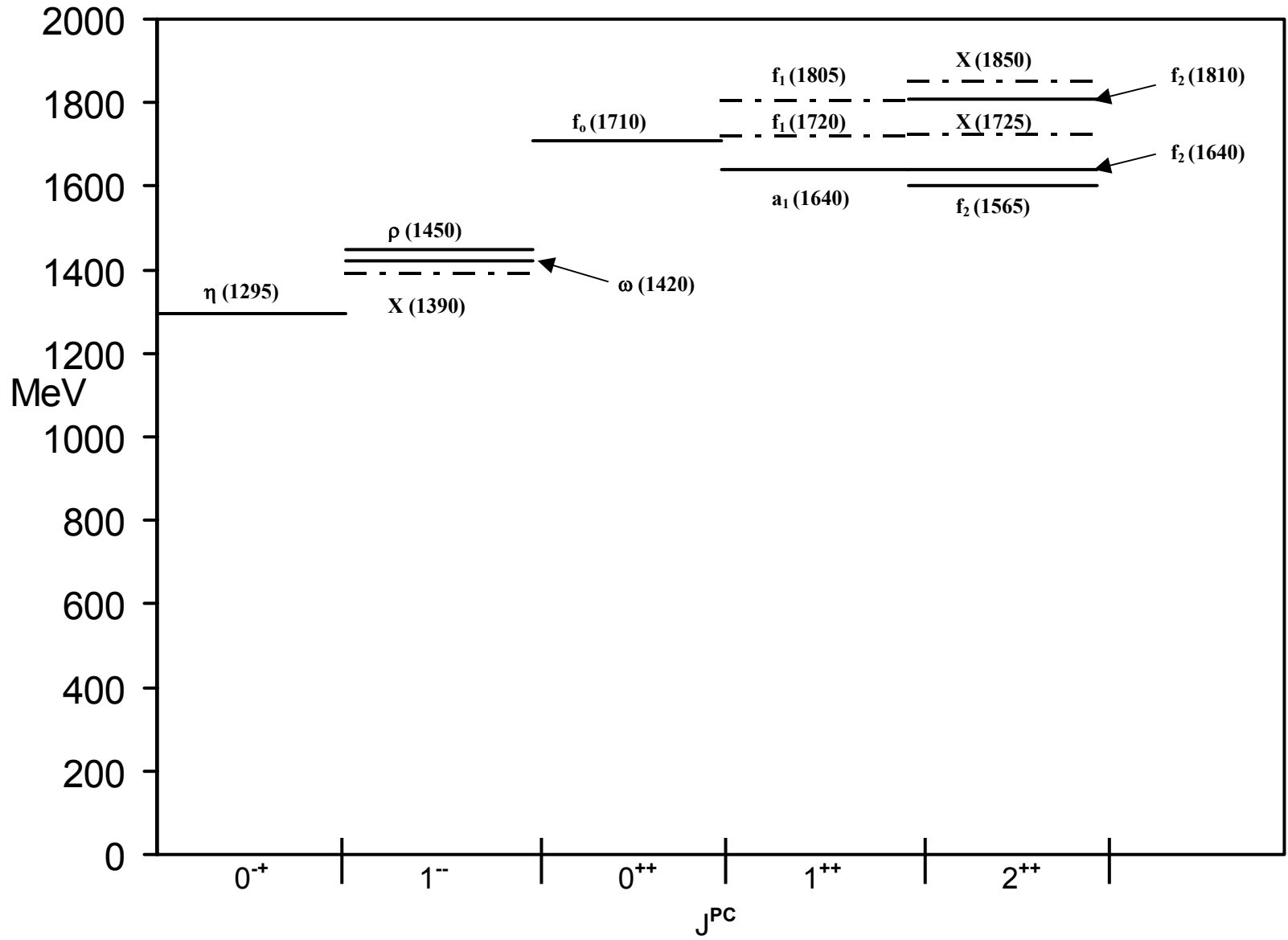

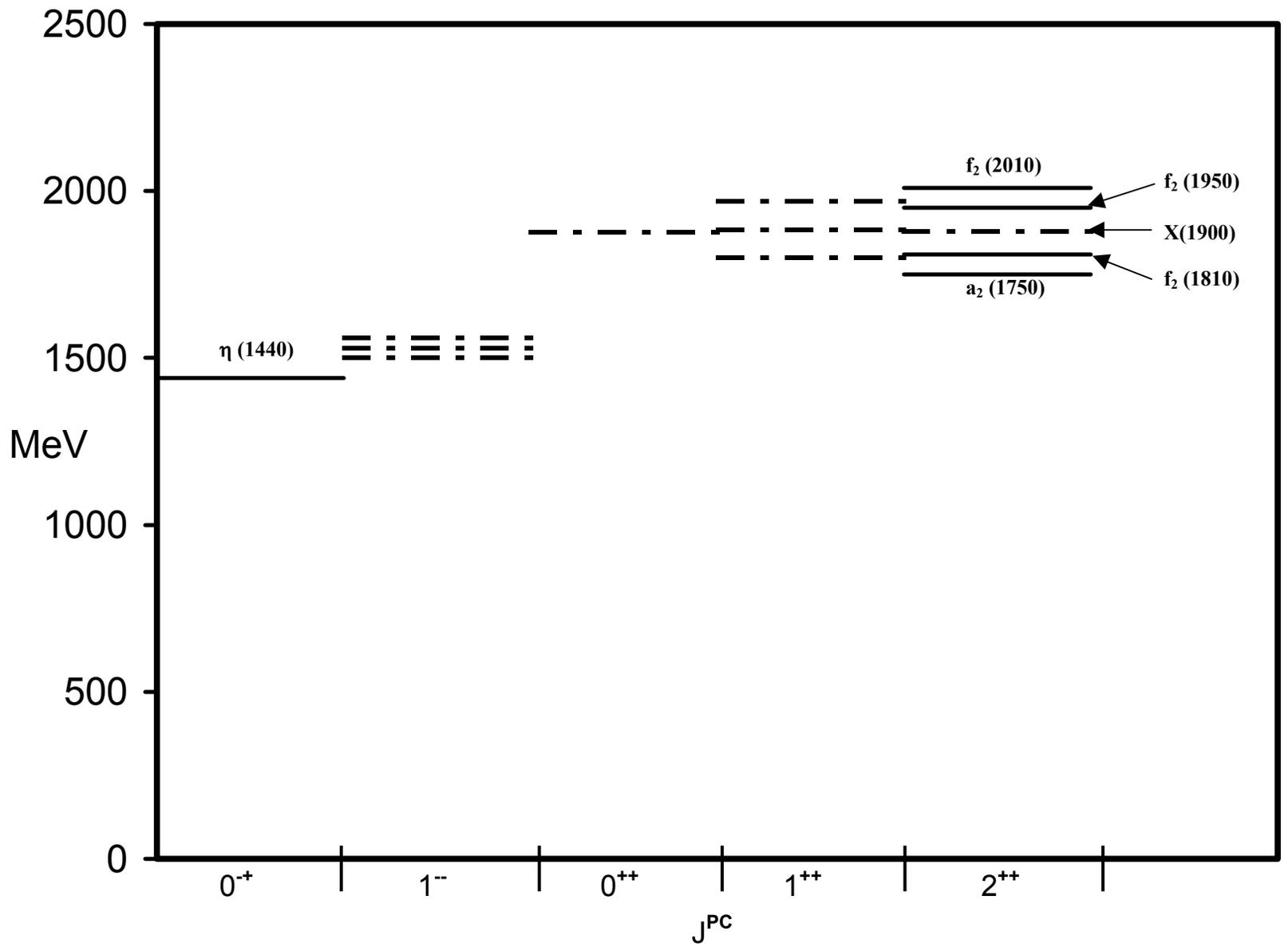

Fig. 10.

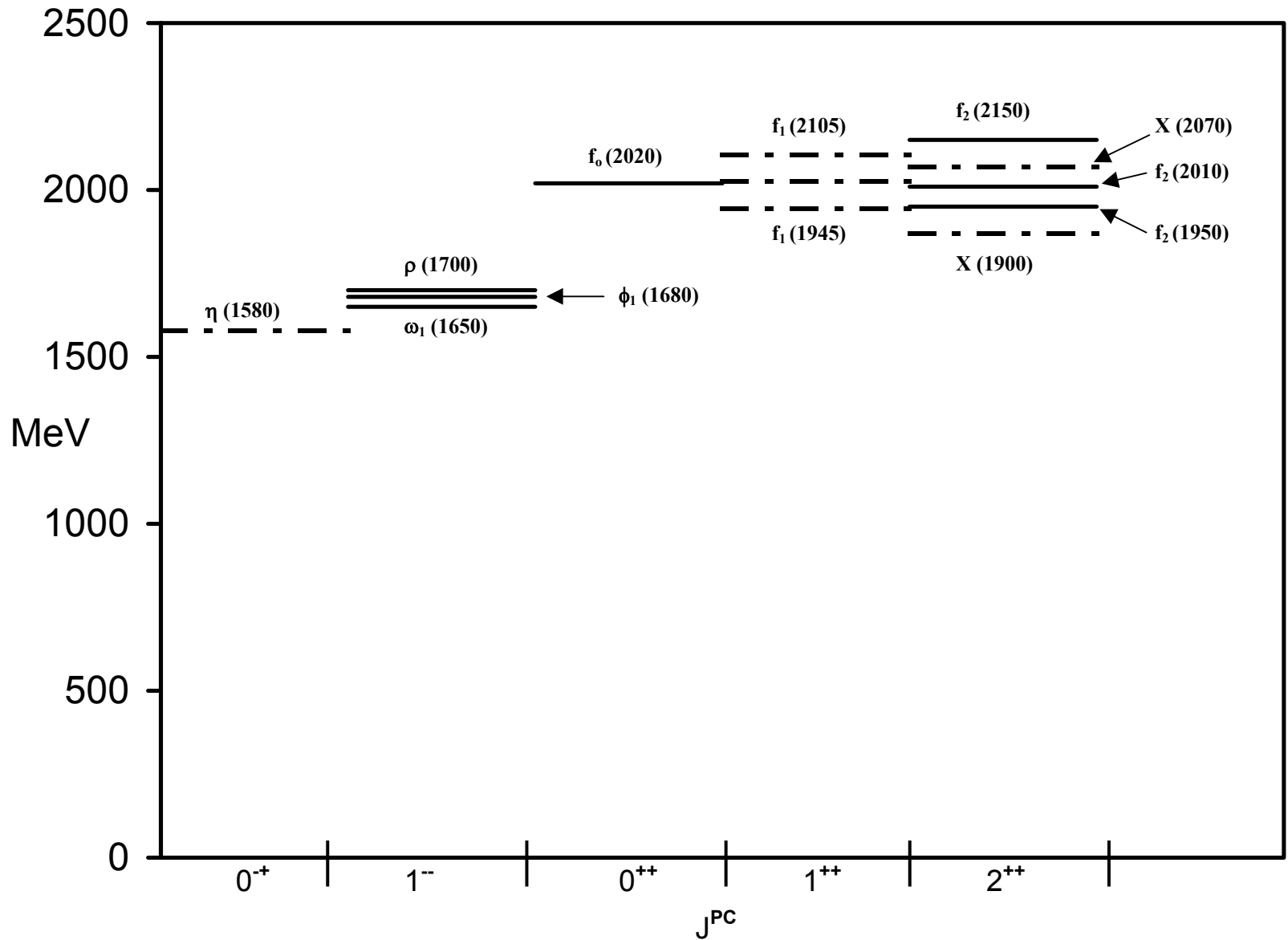

**Fig. 11.**

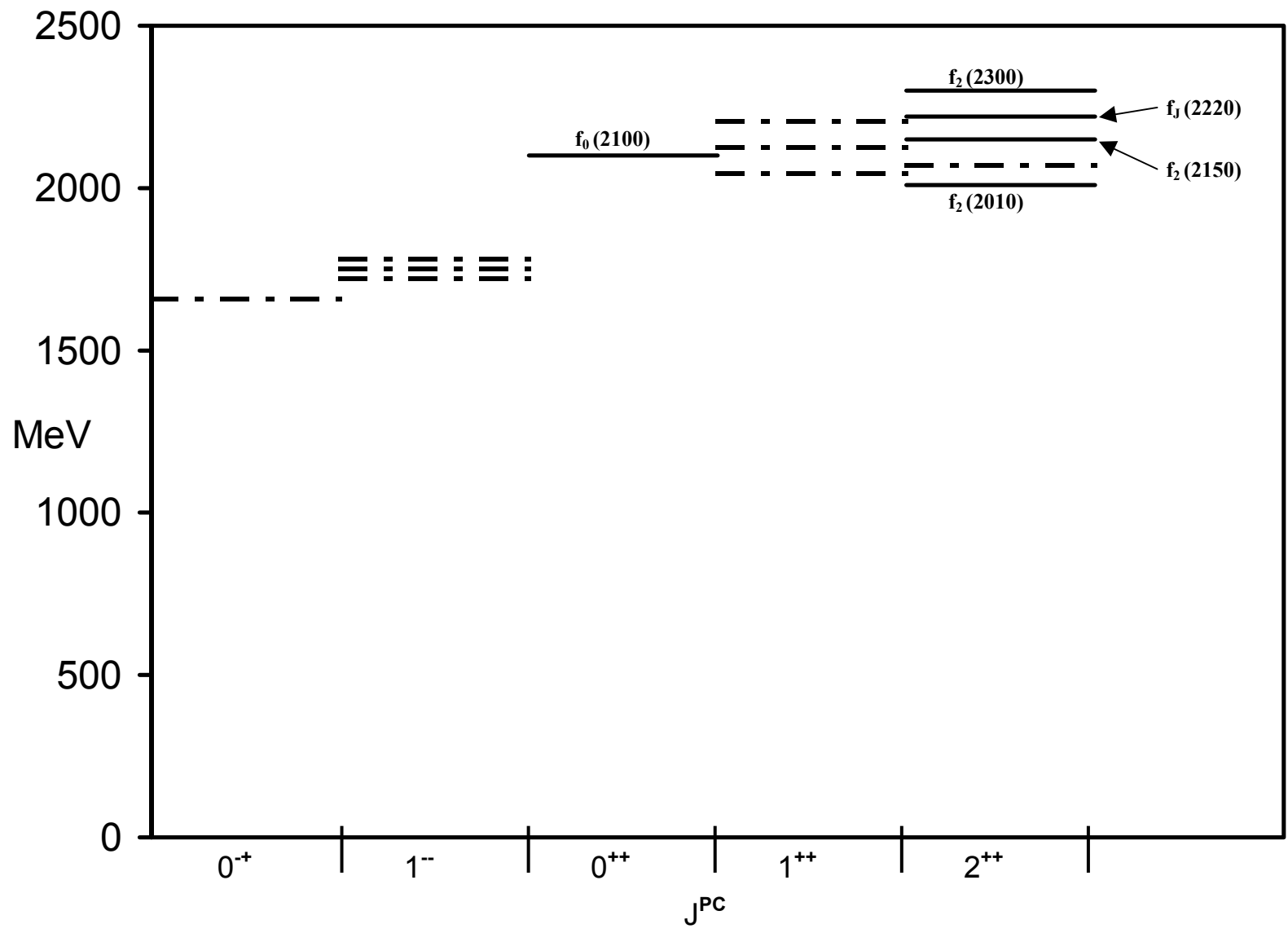

**Fig. 12.**

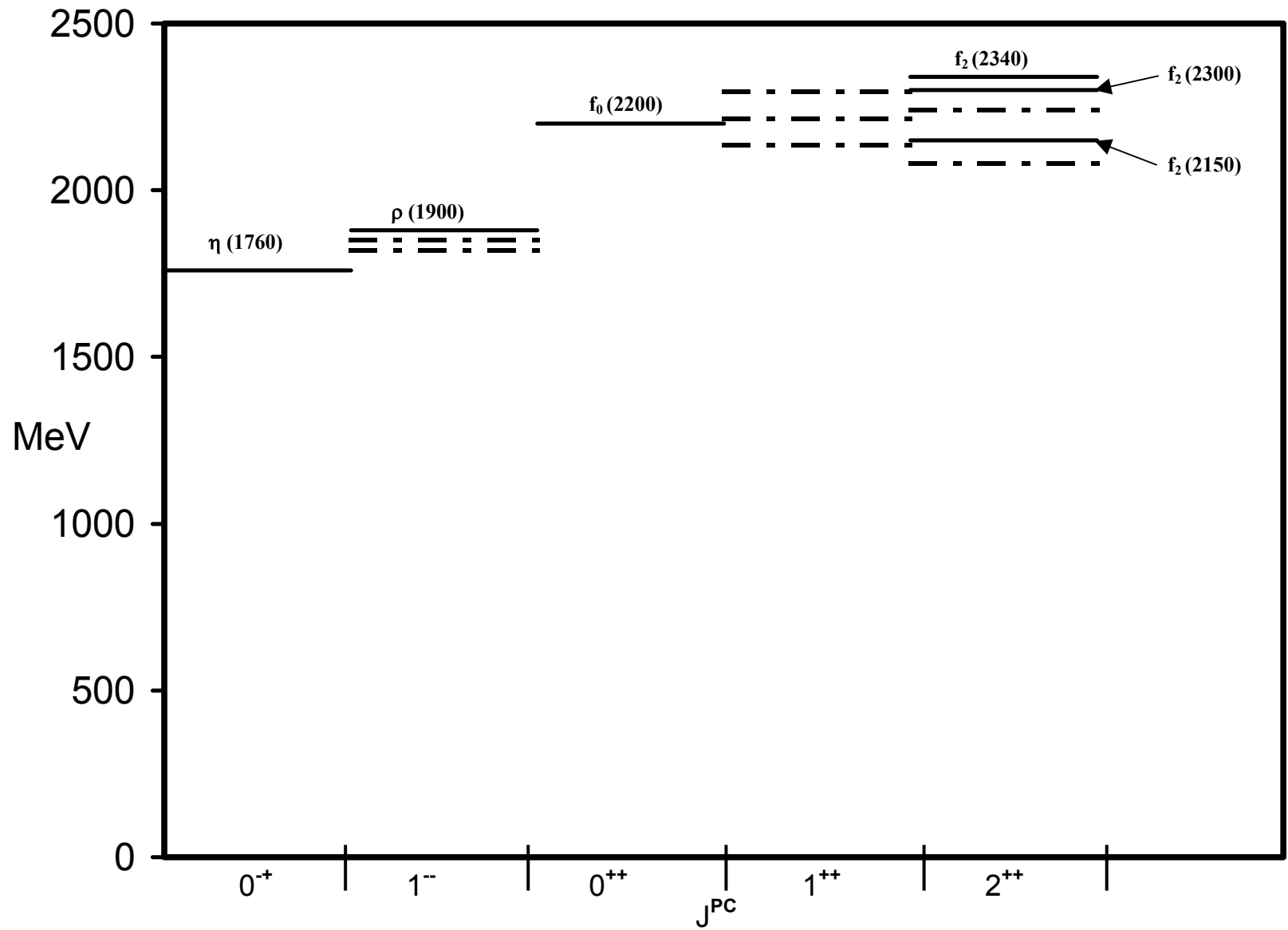

Fig. 13.

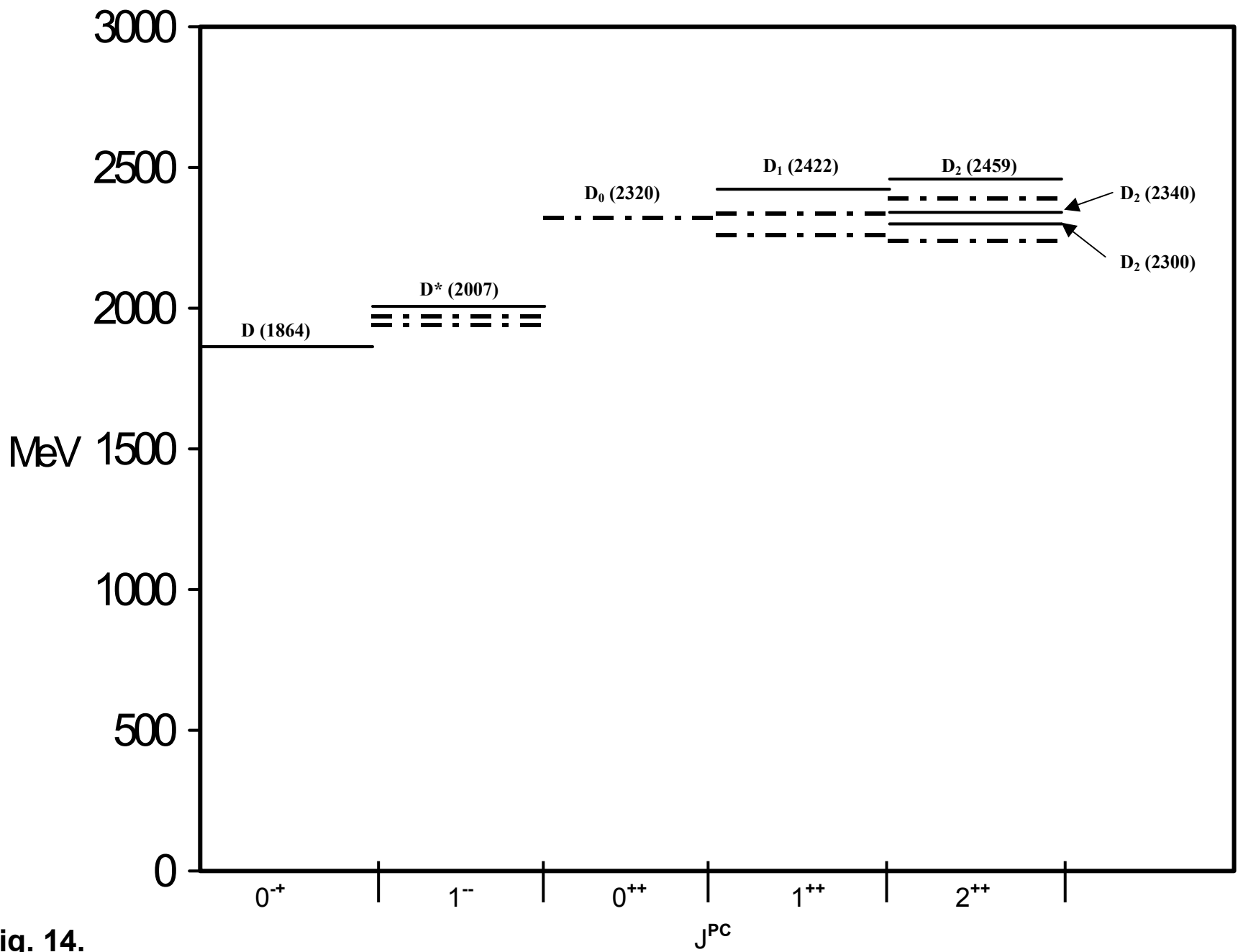

**Fig. 14.**

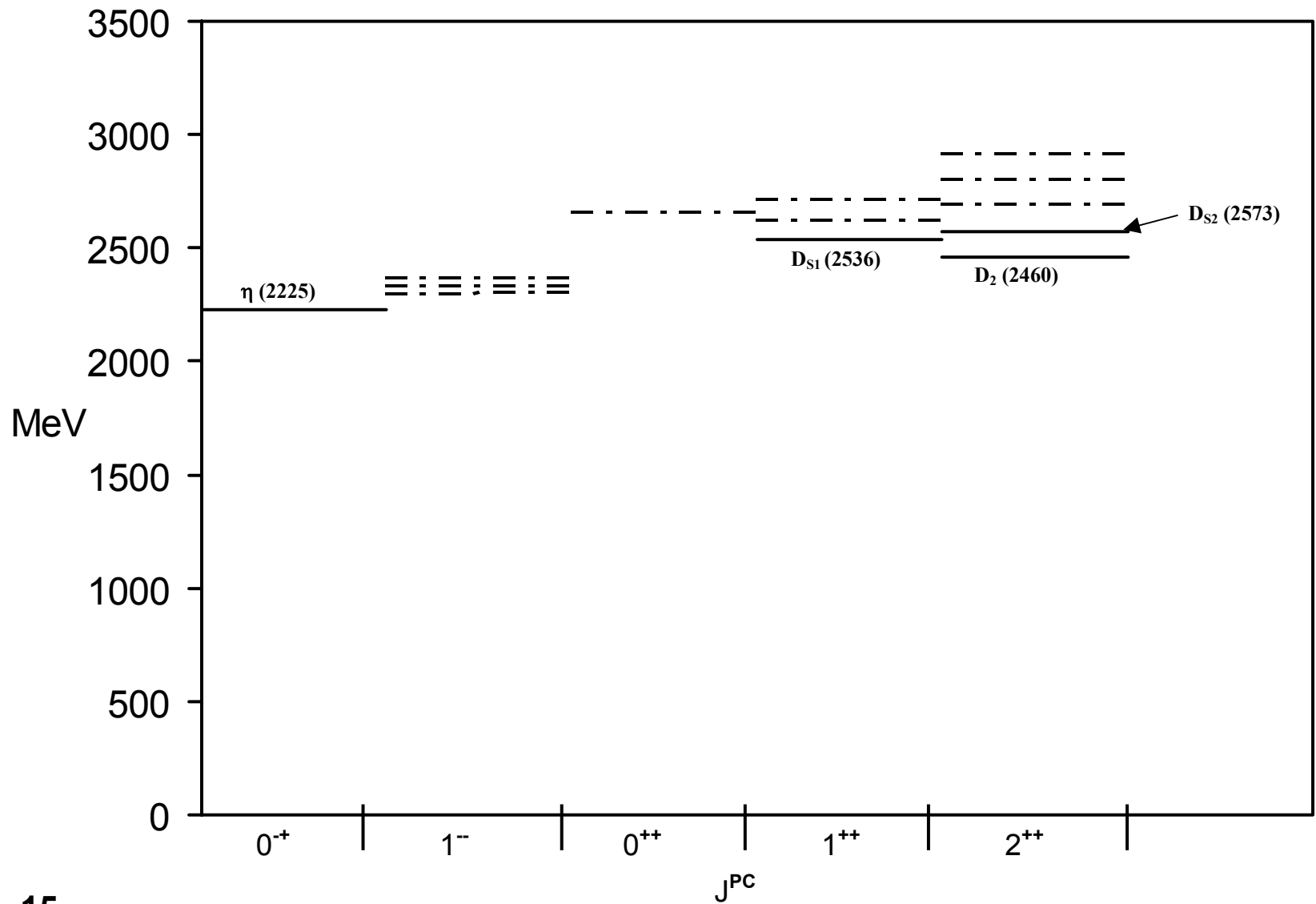

Fig. 15.